\documentclass{aa}
\usepackage{graphicx}
\usepackage{amssymb}
\usepackage{rotating}

\begin{document}
   \title{The Star Cluster Population in the Tidal Tails of NGC 6872
   \thanks{Based on Observations at the Very Large Telescope of the
   European Southern Observatory, Paranal/Chile under Program
   71.B-0596.
 Table~\ref{sources}, containing the photometry,
   positions, and derived properties of each source is available in
   electronic form ec
at the CDS via anonymous ftp to cdsarc.u-strasbg.fr (130.79.128.5)
or via http://cdsweb.u-strasbg.fr/cgi-bin/qcat?J/A+A/}}

   \titlerunning{Star clusters in tidal tails}

   \author{N. Bastian \inst{1,2}, M. Hempel \inst{1}, M. Kissler-Patig
            \inst{1}, N. Homeier \inst{1,3}, and G. Trancho \inst{4,5}   
          }

   \authorrunning{N. Bastian et al.}

   \offprints{bastian@astro.uu.nl}

   \institute{$^1$European Southern Observatory, Karl-Schwarzschild-Strasse 2 
               D-85748 Garching b. M\"{u}nchen, Germany \\
	       $^2$Astronomical Institute, Utrecht University, 
              Princetonplein 5, NL-3584 CC Utrecht, The Netherlands \\
             $^3$ Department of Physics and Astronomy, Johns Hopkins
               University, 3400 North Charles Street, Baltimore, MD
               21218, USA\\
	       $^4$ Gemini Observatory, c/o AURA, Casilla 603, La
               Serena Chile\\
	       $^5$ Universidad de La Laguna, Av. Astrofisico
               Francisco Sanchez sn, 38206, La Laguna, Tenerife, Spain
              }

   \date{Received Oct. 2004; accepted: }


   \abstract{ We present a photometric analysis of the rich star cluster
   population in the tidal tails of NGC~6872.  We find star clusters
   with ages between 1 - 100 Myr distributed in the tidal tails, while
   the tails themselves have an age of less than 150 Myr.  Most of the
   young massive ($10^{4} \le M/M_{\odot} \le 10^{7}$) clusters are
   found in the outer regions of the 
   galactic disk or the tidal tails.  The mass
   distribution of the cluster population can be well described by a
   power-law of the form $N(m) \propto m^{-\alpha}$, where $\alpha =
   1.85 \pm 0.11$, in very good agreement with other young cluster populations
   found in a variety of different environments.  We estimate the star
   formation rate for three separate regions of the galaxy, and find
   that the eastern tail is forming stars at $\sim 2$ times the
   rate of the western tail and $\sim 5$ times the rate of the main
   body of the galaxy.   By comparing our
   observations with published N-body models of the fate of material
   in tidal tails in a galaxy cluster potential, we see that many of
   these young clusters will be lost into the intergalactic medium.  We
   speculate that this mechanism may also be at work in larger galaxy
   clusters such as Fornax, and suggest that the so-called
   ultra-compact dwarf galaxies could be the most massive star
   clusters that have formed in the tidal tails of an ancient galactic
   merger. 
    }      
\maketitle

%

\section{Introduction}

An increasing number of studies are showing the great variety and
abundance of environments that are conducive to the formation of young
massive stellar clusters (YMCs).  From the centers of merging galaxies
(e.g.~Miller et al. 1997; Whitmore et al. 1999) to nuclear starburst
rings (Maoz et al. 2001) and even in normal spiral galaxies (Larsen \& 
Richtler 2000).  In addition, recent studies have shown that YMCs
can form outside the main body of a galaxy, in tidal tails (Tran et
al. 2001, Knierman et al. 2003) and between galaxies in compact
groups (Gallagher et al. 2001).  Indeed, it appears that wherever the
local star-formation rate is high enough, YMCs are certain to be
present.

The goal of this study is to determine the properties of the rich star
cluster population in the tidal tails of the NGC~6872.  By studying
the age distribution of the clusters with respect to the spatial
distribution, we can gain insight into the cluster formation process on
large ($\sim$galaxy) scales.  Additionally, we shall compare the
derived parameters of the population as a whole with that of more
typical cluster populations in order to search for differences related
to environmental influence.

Recently, Knierman et al. (2003) have surveyed the tidal tails of four
mergers of spiral galaxies.  They found that only one of the six tails 
studied (the west tail of NGC~3256) showed a large number of star
clusters, while three others 
showed evidence of a small number of clusters.  The authors suggest that the
presence of tidal dwarf galaxies may hamper the formation of YMCs in
the rest of the tidal tail.  In this sense, NGC~6872 is akin to
NGC~3256, as it lacks a clear tidal dwarf galaxy in each of its
tails. 

\begin{figure}
 \begin{center}
     \includegraphics[width=8cm]{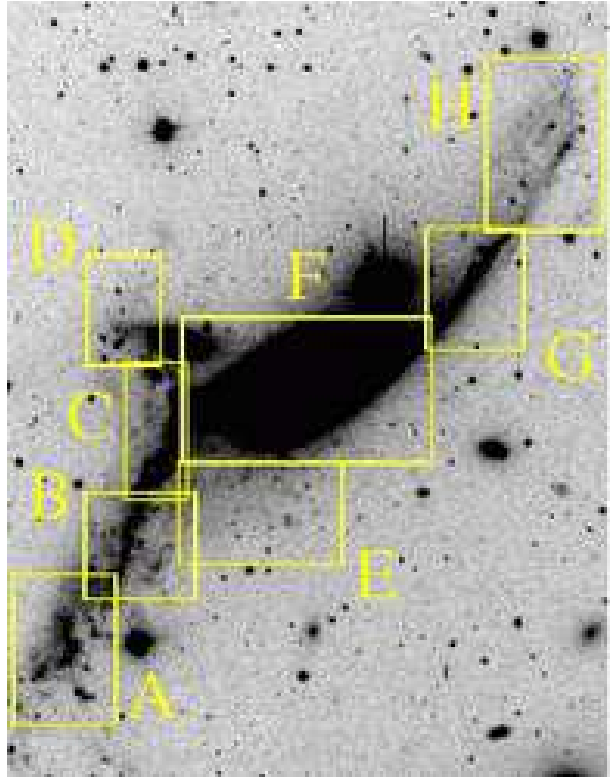}
      \caption{B-band image of NGC~6872.  Regions of interest
 (i.e.~high YMC density) are
 labeled.  North is to left, and east is to the bottom.  The image is
 297$''$ by 382$''$ or $\sim$18.7 kpc by 24.1 kpc.}
      \end{center}
         \label{image}
   \end{figure}

\begin{figure}
  \includegraphics[width=8cm]{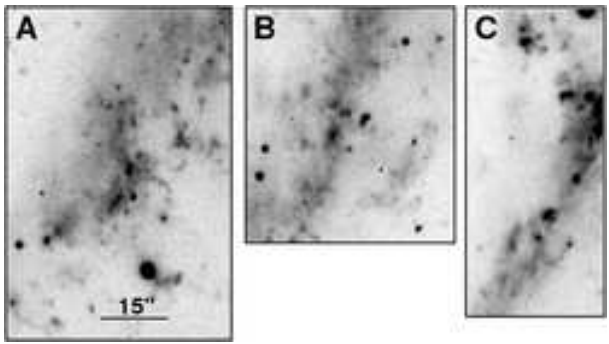}
  \caption{B-band images of the individual regions A, B, and C in the
  east tidal tail.  All images have
  the same scale.}
  \label{regions_abc}
\end{figure}

\begin{figure}
  \includegraphics[width=8.1cm]{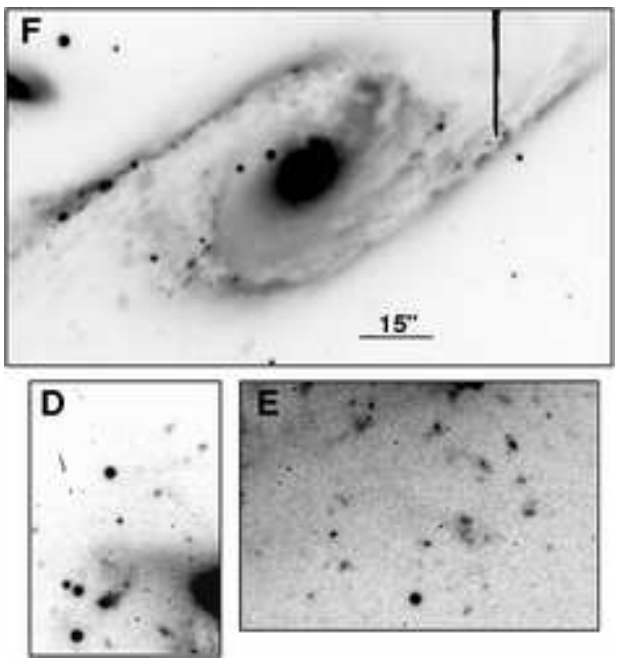}
  \caption{B-band images of the regions D, E and F in the main body of
  the galaxy and the region between the two galaxies.  All images have
  the same scale.}
  \label{regions_dfe}
\end{figure}

\begin{figure}
  \includegraphics[width=8cm]{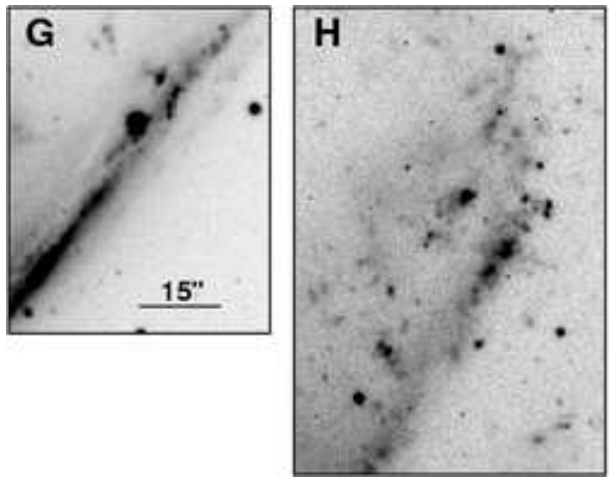}
  \caption{B-band images of the individual regions G and H in the west
  tail.  All images have the same scale.}
  \label{regions_gh}
\end{figure}

NGC~6872 is a barred spiral galaxy that is interacting
with IC~4970, a smaller S0 galaxy (de Vaucouleurs et al. 1991), which
is the elliptical structure in the center left of Fig.~\ref{image}.  In
Figure~\ref{image} an optical ({\it B}-band) image of NGC~6872 is shown, and 
each of the regions labeled and shown in
more detail in the subsequent figures
(Figs.~\ref{regions_abc},~\ref{regions_dfe}, \&~\ref{regions_gh}).
Note that all panels are based on the same scaling (see panel A, F, and
G).  

The NGC~6872/IC~4970 interaction has been modeled by
Mihos et al. (1993) and Horellou \& Koribalski (2003)
who both find a 
mass ratio between NGC~6872 and IC~4970 of $\sim$ 5:1, and a time since
periapse of $\sim 145$ Myr, i.e.~the event that triggered
the formation of the tails.  Two long, thin tidal tails extend from
the disk of NGC~6872, the optical component of which terminates some 60 kpc
from the galactic center.  NGC~6872 has a rich star cluster population
in its tidal tails, first 
noted by Horellou \& Koribalski (2003).  The combination of new U-band
imaging, archival B, V, and I imaging, and Fabry-Perot $H\alpha$ data
taken from the literature allow an analysis of the
cluster population as a whole including their formation history and
spatial position within the tails.  This in turn will provide insight
into the formation and evolution of star clusters in the special
circumstance of galaxy interactions.

This study is set up in the following way.  In
\S~\ref{obs} we present the observations and the methods of reduction
and photometry employed.  In \S~\ref{clusterpop} we study the colour
distribution of the cluster population and infer each cluster's age
and mass.  The youngest clusters are discussed in
\S~\ref{youngclusters} which is followed by an estimation of the star
formation rate in the individual regions of the galaxy in
\S~\ref{specu}.  Our observational results are combined with models of
galaxy interactions
from the literature in order to explore the formation and future
evolution of the cluster population in \S~\ref{modelcomp}, and in
\S~\ref{conclusions} we summarize our results.  
 
The heliocentric redshift of NGC~6872 is 4701 km/s (de Vaucouleurs et
al. 1991), which, assuming
$H_{0}$  = 72 ${\rm km} {\rm s}^{-1} {\rm Mpc}^{-1}$, corresponds to a
distance of $\sim 65$ Mpc and 
a distance modulus of 33.9 mag.

\section{Observations, reduction and photometry}
\label{obs}
To derive the ages and masses of extragalactic YMC using photometry,
multi-band observations spanning a wide range in wavelength are
required (e.g.~de Grijs et al. 2003a).  To this end, we retrieved
images taken with {\it VLT-FORS1} of NGC~6872 in the {\it B}, {\it V},
and {\it I} bands from the {\it ESO/VLT} archive.  The {\it B} band
image was a single 
exposure of 600s, while the {\it V} and {\it I} band images consisted
of two exposures each of 150s and 200s respectively.  Additionally,
$H{\alpha}$ Fabry-Perot observations of NGC~6872 were kindly provided
by C.J.~Mihos and used to corroborate our results (see Mihos et
al. 1993 for a description of the observations).   

As shown by Anders et al. (2004) {\it U}-band
observations are essential for photometric investigations of young
star clusters.  We have therefore complemented
the existing archival data with {\it VLT-FORS1} {\it U-}band observations,
taken on May 6th,
2004.  Three {\it U} band images of 300s each were taken.  All images
were flat-fielded and the bias subtracted in the standard way.

Sources were identified using the {\it DAOFIND} routine in {\it
IRAF}.  Aperture photometry was performed with an aperture,
inner background radius, and outer background radius of 5/7/9
pixels (1 pixel$ = 0.2''$).  Aperture corrections to a 50 pixel (10$''$) radius
were calculated 
using bright isolated sources.  The applied corrections were $-0.71$,
$-0.21$, $-0.16$, $-0.16$ mag for the {\it U, B, V, {\rm and}, I}
respectively.  The large difference between the aperture correction for
the {\it U} band relative to the other bands was due to the
differences in seeing.  The seeing during the {\it B, V, {\rm and} I}
observations 
was $\sim~0.55$ arc seconds, while for the {\it U} band observations
it was $\sim~1.5$ arc seconds.  Photometric zero points were
determined using Landolt (1992) standard fields. 

Completeness limits were determined by adding artificial sources with
magnitudes between 21 and 26 mag, using 
the {\it IRAF} task {\it ADDSTAR}.  The sources and their magnitudes
were found following the procedure described above.  The 90\% completeness
limits are 23.75, 24.0, 23.5, and 22.75 for the U, B, V, and I bands
respectively.  We do not expect our sample to be contaminated by individual
luminous stars in NGC~6872, as our completeness limit corresponds to
M$_{V}$ = -10.4 (at the distance of NGC~6872) which is almost a  magnitude and
a half brighter than the {\it cluster selection} limit used by
Whitmore et al. (1999).  Foreground stars, however, may contaminate
our sample, and will need to be accounted for.

We have corrected all observed sources for Galactic extinction
($A_{V} = 0.15$, Schlegel, Finkbeiner, \& Davis 1998).

\section{The cluster population}
\label{clusterpop}
\subsection{Colours}

\begin{figure}[!h]
\includegraphics[width=7.5cm]{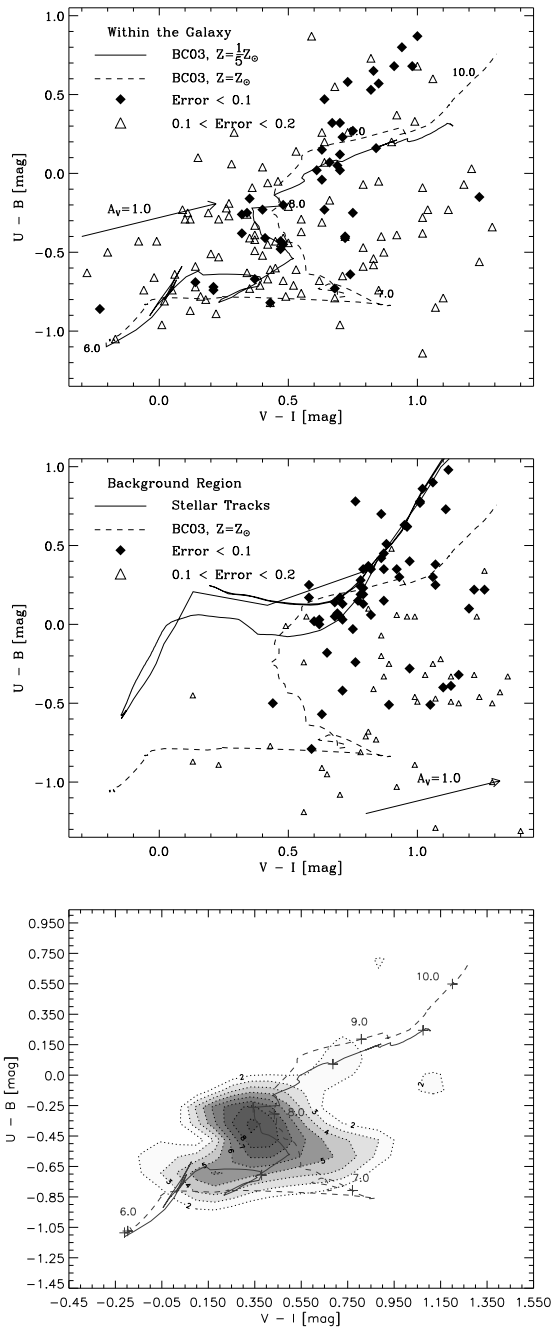}
      \caption{{\bf Top:} Colour-colour diagram for sources in NGC~6872,
     including the tidal tails.  Galactic extinction has been
     removed.  The extinction correction vector of $A_{V} = 1$ is
     shown. Here error refers to the error in the measured colours.
     SSP model tracks are over-plotted (see text).  The numbers along
     the model tracks refer to the logarithm of the 
     age in years.  {\bf Middle:} Colour-colour diagram for the
     background region in the images. Here the solid line is the
     stellar model tracks (see text for a description) 
     {\bf Bottom:}  The contour levels of the middle figure subtracted from
     the top figure (see text for details).  Model tracks are the same
     as in the top panel. }
         \label{colourplot}
   \end{figure}

Figure~\ref{colourplot} shows the colour-colour ($V-I$ vs $U-B$) diagram
for point-like sources in both tidal tails as well as in the
central region of NGC~6872.  The solid and dashed lines are the SSP
model tracks (Bruzual \& Charlot 1993 - 2000 version) for $\frac{1}{5}
Z_{\odot}$
and $1 Z_{\odot}$ respectively, and salpeter IMF.   For comparison we
also constructed the 
colour-colour diagram for the background sample (sources outside the
labeled areas in Fig.~\ref{image}), which is shown in the middle panel
of Fig.~\ref{colourplot}.  Presumably the sources in the background
field are foreground
stars and background galaxies.  This is shown in the middle panel of
Fig.~\ref{colourplot} by also showing the stellar model tracks (Z =
0.008, $0.8 \le M/M_{\odot} \le 5.2$, 100 Myr - Lejune \& Schaerer
2001).  Most of the bright sources 
follow the stellar tracks quite nicely (the small offset to the blue
is caused by the correction for Galactic extinction).

We then  removed the contamination from our sample by subtracting
the 'background' field from the 'galaxy' field (normalizing
to equal areas).  This was done statistically by dividing the
colour-colour plane for both the background and galaxy fields, into
square regions ($\Delta(U - B)$ \& $\Delta(V - I) = 0.3$), and then
summing the number of sources within each 
colour box.  The resulting background grid was then subtracted from
the galaxy grid, and the result is shown as a contour plot in
the bottom of Fig.~\ref{colourplot}.

There is a clear over-density of very blue objects, with colours
consistent with ages between 1 - 100 Myr, 
peaked at a few Myr (following the extinction vector), as shown by the
contour plot (bottom panel in 
Fig.~\ref{colourplot}).
The tails have ages of $\sim 145$ Myr (Mihos et al. 1993), and
many of the very young clusters are found at a large distance from the
main body of the galaxy.  Hence
these structures must have formed within the tail, and were not formed
inside the main body of the galaxy and ejected into the tail.

\subsection{Ages}

In order to more quantitatively investigate the properties of these
clusters, we attempt to derive the age, mass, and extinction of
each cluster.  To do this we employ the three dimensional spectral
energy fitting algorithm ({\it 3DEF method}) first proposed by Bik et
al. (2003).  A detailed description of the method is given there, and
as such we shall only provide a brief summary here.  The {\it 3DEF}
method compares each cluster's observed spectral energy distribution
({\it SED}) to that of a grid of simple stellar population ({\it
SSP}) models of ages between 1 Myr and 11 Gyr, and extinctions ranging
from $0
\le A_{\rm V} \le 4$ in equal steps of 0.02.  We adopt the models of
Bruzual \& Charlot (1993 - 2000 version) with a metallicity of
       $\frac{1}{5} Z_{\odot}$.  The metallicity 
       was chosen to most nearly match the observed metallicity of
       tidal dwarf galaxies ($Z \sim \frac{1}{3} Z_{\odot}$ - Duc \&
       Mirabel 1998), which are objects which also form in the tidal
       debris of merging galaxies.  For a detailed description of the
       effects of an apriori metallicity assumption, see Anders et
       al. (2004).  Here we note that if solar metallicity is used
       the derived ages become less, on average, and therefore so do
       the determined masses.  We will discuss how our assumed
       metallicity affects our results where applicable.

\begin{figure*} 
\begin{centering}
    \includegraphics[width=16cm]{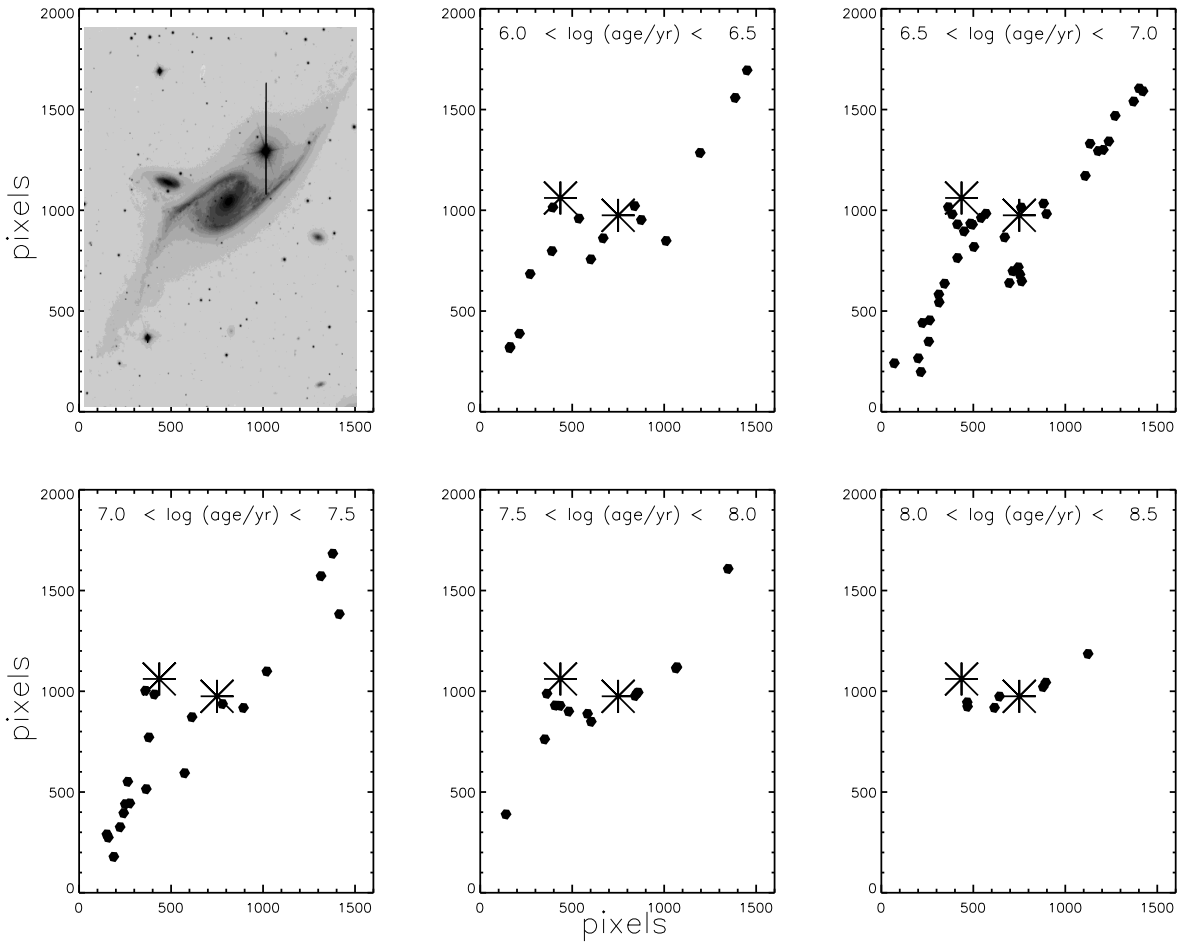}
     \caption{The age of the detected clusters as a function of their
     position.  The x and y axis are in pixels (1 pixel = 0.2 arc
     seconds = 63 pc).  The large asterisks in the figure mark the center
     of NGC~6872 and IC~4970.}   
         \label{posage}
	 \end{centering} 
   \end{figure*}

\begin{figure*}
\begin{centering}
    \includegraphics[width=16cm]{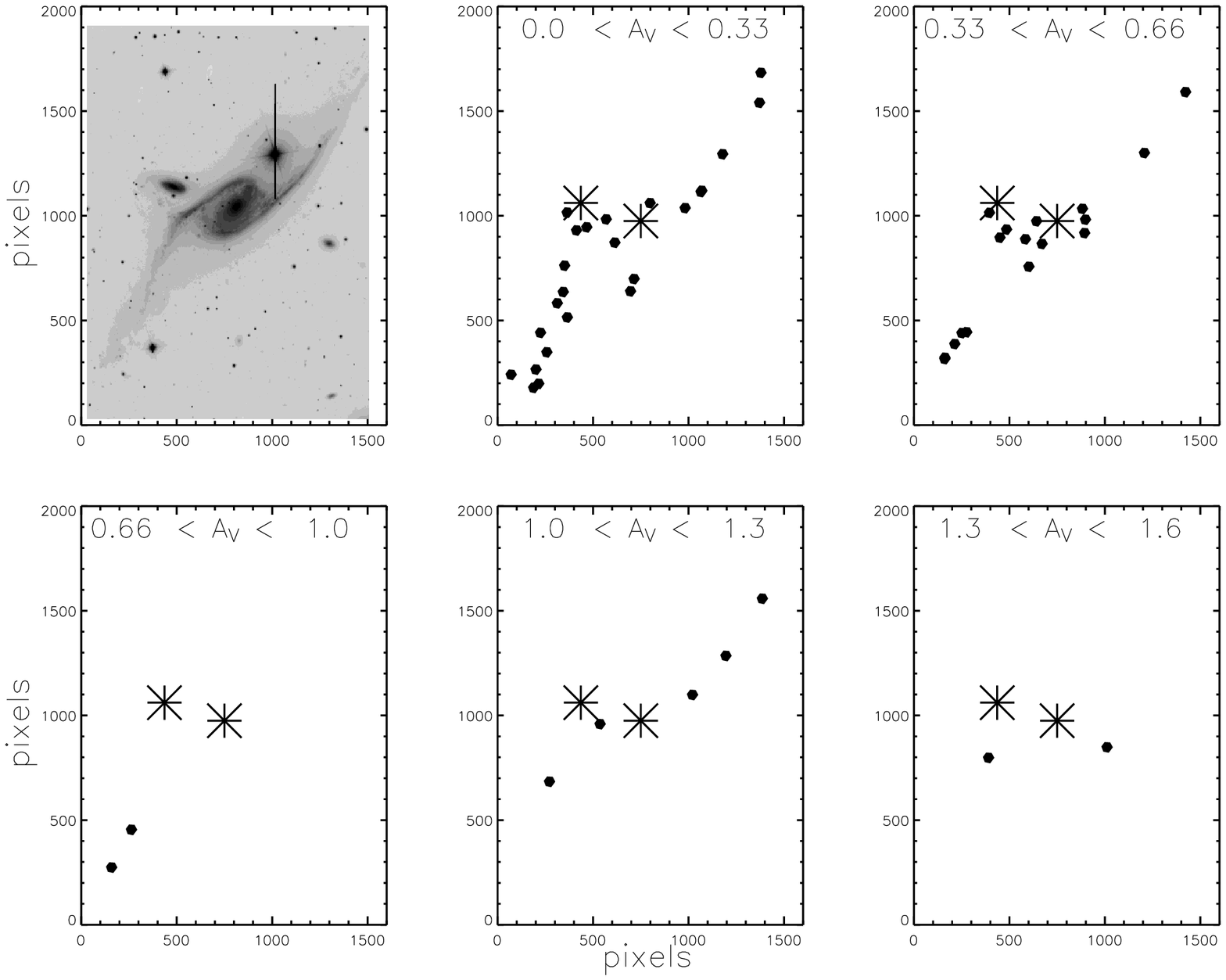}
     \caption{The derived extinction of the detected clusters as a
     function of their
     position.  The x and y axis are in pixels (1 pixel = 0.2 arc
     seconds = 63 pc).  The large asterisks in the figure mark the center
     of NGC~6872 and IC~4970.}
         \label{pose}
	 \end{centering} 
   \end{figure*}

Applying this technique to the clusters in NGC~6872, we derive ages
between a few Myr to a few hundred Myr.  No clusters older than
145 Myr (the age of the interaction) are found
inside the tidal tails.  Figure~\ref{posage} shows the derived ages
(divided into five age bins) as a function of position in the galaxy.
From this we conclude that the youngest clusters are spread throughout
the tidal tails, while the older clusters are concentrated towards the
main body of the galaxy.  Some caution must be taken in the
interpretation of this figure (also Figs.~\ref{pose}~\&~\ref{posmass}) as
foreground stars and background galaxies may be contaminating the
sample.  We have attempted to eliminate contaminating sources by
applying a reduced $\chi^2$ criterion to the fits (Bik et al. 2003).
For this we adopted the value from Bik et 
al. (2003) of a maximum $\chi_{\nu}^2$ of 3 between the observed
cluster spectral energy distribution and that of the best fitting
model.  Adopting solar metallicity for the SSP model does not change
the general properties of the spatial distribution, except for a
larger number of sources with ages between 30 Myr and 100 Myr in the
tidal tails.

From Fig.~\ref{posage} we note that clusters with ages less
than $\sim 30$ Myr (log (age/yr) = $\sim 7.5$) are spread throughout
the galaxy, including the tidal tails.  Clusters older than this are
concentrated towards the center of the galaxy.  This can be
interpreted as a delay between the merger event which caused the
tidal tails, and cluster formation inside the tidal tails.

Additionally, the {\it 3DEF} method solves for the extinction for each
cluster simultaneously with the age.  Most of the clusters in
this study were fit with low extinction values ($ 0 \le A_{V} \le 0.5$
- after correction for foreground extinction).  Figure~\ref{pose}
shows the position in the galaxy for five different extinction bins.
We do not find any strong correlation between position and extinction.
In particular no increase is found in the average extinction
along the tidal tails from HI gas poor to gas rich regions.  However, the
two groups of star clusters far into the tidal tails in the $0.5 <
A_{V} < 1.0$ bin seem to be associated with peaks in the HI intensity map
of Horellou \& Koribalski (2003).

\subsection{Masses}

At a distance of $\sim 65$ Mpc,
one {\it FORS1} pixel corresponds to 63 pc, which leaves room to discuss to
which degree our cluster detections might not refer to single clusters
but to multiple, unresolved clusters.  Given that young
clusters seem to be themselves clustered (e.g.~Larsen 2004) it is likely
that some of our sources are made up of multiple clusters.  This may
cause us to over-estimate the mass of the observed clusters, as the
mass that we measure is in fact the total mass of multiple clusters.
Therefore our mass estimates for some sources represent the total mass 
of a complex of clusters, which, given the scale of the resolution ($\lesssim
150$ pc), implies that these clusters formed within the same star forming
complex.  High resolution imaging is required to estimate to what degree our
mass estimates are affected by this effect.  Note, however, that this
effect will have a negligible influence on our age and extinction results.

\begin{figure*}
\begin{centering}
    \includegraphics[width=16cm]{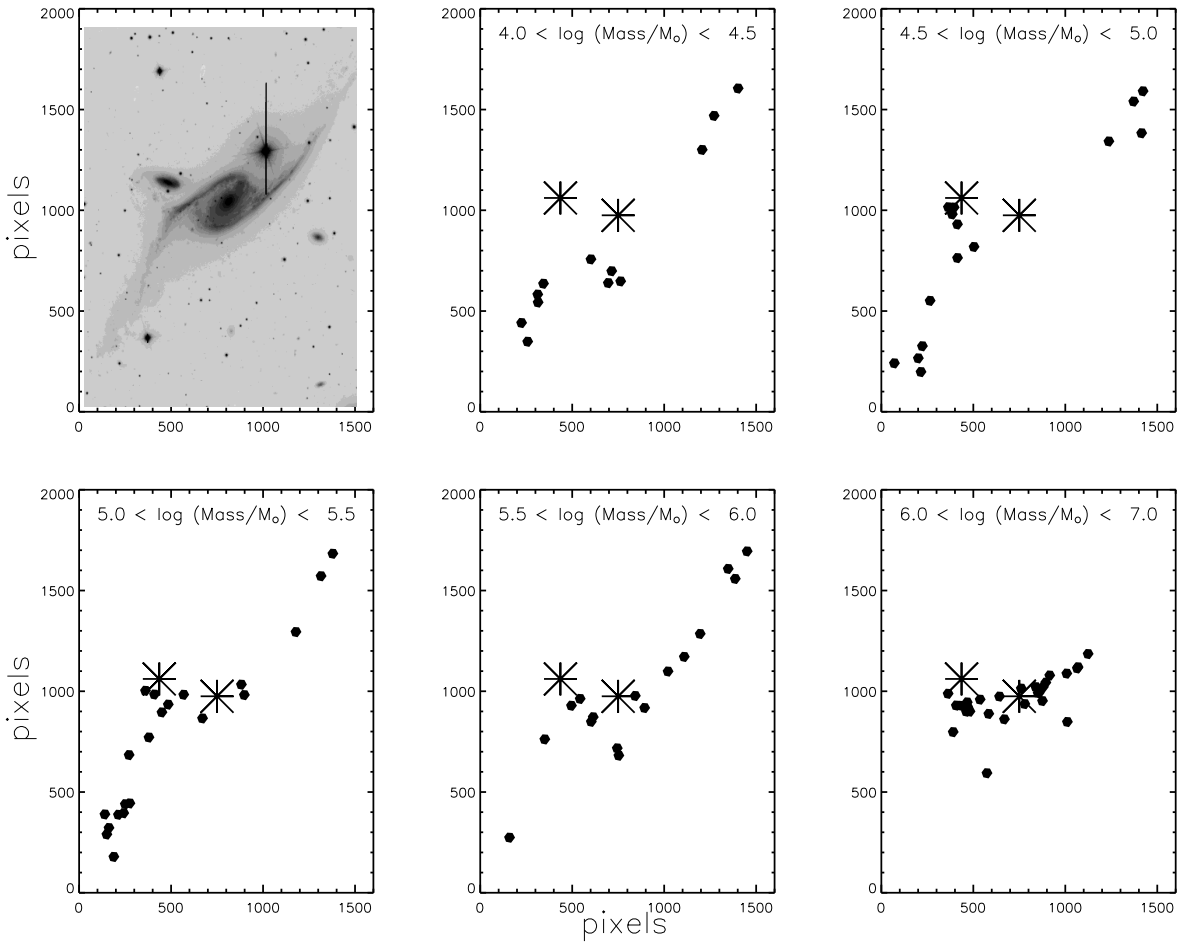}
     \caption{The mass of the detected clusters as a function of their
     position.  The x and y axis are in pixels (1 pixel = 0.2 arc
     seconds = 63 pc).  The large asterisks in the figure mark the center
     of NGC~6872 and IC~4970.}   
         \label{posmass}
	 \end{centering} 
   \end{figure*}

Figure~\ref{posmass} shows the position of the sources inside the
galaxy and tails in five different mass bins.  As in
Fig.~\ref{posage} we have attempted to remove foreground stellar
contamination by applying a $\chi^{2}_{\nu}$ criterion.  There are
clusters with masses up to $10^{6} M_{\odot}$ forming in the tails,
and clusters with even larger masses forming inside the main
galaxies and region between the two interacting galaxies.  The
assumption of solar
metallicity for the template SSP models does not change the
qualitative description of the spatial positions of the clusters for
different mass ranges given above.

Although we are limited to rather low spatial resolution and are
potentially biased towards identifying multiple, closely spaced
clusters as a single massive clusters (see above), we move forward, and
construct a mass distribution for the clusters in our sample.  The mass of each
cluster can be estimated by combining the derived age and extinction
of each cluster, the measured brightness, the known distance modulus,
and the predicted mass-to-light ratios of SSP models.
Figure~\ref{n6872-mass} shows the 
derived mass distribution for the young clusters ($\le 10$ Myr) in our
sample.  The assumed metallicity has almost no influence on the mass
distribution.  The dashed vertical line in the figure
is the detection limit of our sample assuming an age of 10 Myr and no
extinction.  Presumably the mass function 
extends to much lower masses, and the observed turnover is simply
caused by the detection limit.  There is, however, a significant
number of high mass ($> 5 \times 10^{5} M_{\odot}$) clusters.  The
shaded histogram is the mass distribution of the young clusters which
were selected based on their H$\alpha$ flux.  This subsample will be
discussed in more detail in \S~\ref{youngclusters}.

\subsubsection{Cluster initial mass function}
\label{clf}

The mass function of young star cluster populations in many
diverse environments can be well fit with a power-law of the
form $N(m) \propto m^{-\alpha}$, where $\alpha = 1.95 \pm  0.26$
(data taken from Table~2 of de Grijs et al. 2003b\footnote{Note
that the exponents of mass functions and luminosity functions have
been combined here.  If the cluster population is close to a single
aged population then $\alpha$ is the same for the mass and luminosity
functions.   For a detailed analysis of the luminosity function of
young cluster systems see Gieles et al. (2005).}).  The error of 0.26 is calculated as the standard deviation from
the mean.  It is curious that this relation holds across 
many orders of magnitude and many different environments, such as the
centers of galaxy mergers with strong starbursts or quiescent disk
galaxies.  This hints at a uniformity in the cluster formation
process, independent of environment.  

As tidal tails in galaxy mergers are an  unexplored
environment for this characteristic we derive the mass function
for the detected clusters in our sample.  We assume a fit of the form
above, with $\alpha$ as the free parameter.  The
results of this procedure are shown in Fig.~\ref{n6872-mass}.  We find
that $\alpha = 1.85 \pm 0.11$, which is somewhat shallow for a young
cluster population, but still within the standard deviation given above.  The
shallowness of the slope
may be caused by crowding problems, as we 
may mistake a tight collection of clusters as a single cluster, hence
overestimating the number of bright clusters and underestimating the
number of faint clusters.   Thus,
we would expect that the cluster luminosity function would become
somewhat steeper with higher resolution images.  Even with this
caveat it is clear that the population within NGC~6872 is
very similar to other young cluster populations in terms of its
mass function (see also review by Whitmore 2002).

\section{The Youngest Clusters}
\label{youngclusters}
In order to test the reliability of our age estimates, we have compared
the spatial distribution of the youngest ($< 10$ Myr) clusters
with the  H$\alpha$ emission Fabry-Perot intensity map of Mihos et
al. (1993), which traces the massive stars in the galaxy.
As shown in Fig.~\ref{ngc6872-ha}, 
the $H\alpha$ intensity (dark indicating the strongest emission),
and the positions of the young ($\lesssim 10$ Myr) massive ($ >
10^{4} M_{\odot}$) 
clusters, coincide.  At almost every maxima of the $H\alpha$ intensity
we find a young massive cluster.  Strong H$\alpha$ emission is
evidence of a very young ($< 8$ Myr) population of massive stars
(e.g.~Whitmore \& Zhang 2002).  Our results, based on colour fitting,
are consistent with the $H\alpha$ emission.

\begin{figure}
     \includegraphics[width=8.5cm]{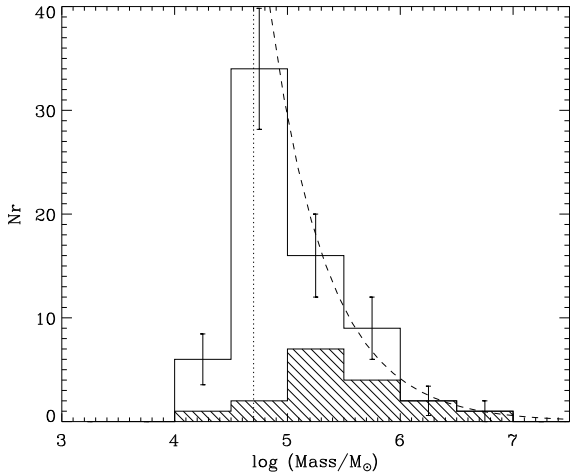}
     \caption{The mass distribution for the young ($<$ 10 Myr)
     clusters.  The shaded histogram are clusters that are
     associated with peaks in the 
     $H\alpha$ intensity image.  The dotted vertical line is the
     completeness limit in the V-band for a 10 Myr old cluster.  The
     dashed line is a fit to the data above the completeness limit of
     the form $N(m) \propto m^{-\alpha}$ with $\alpha = 1.85$.}  
         \label{n6872-mass}
   \end{figure}

\begin{figure}
     \includegraphics[width=8cm]{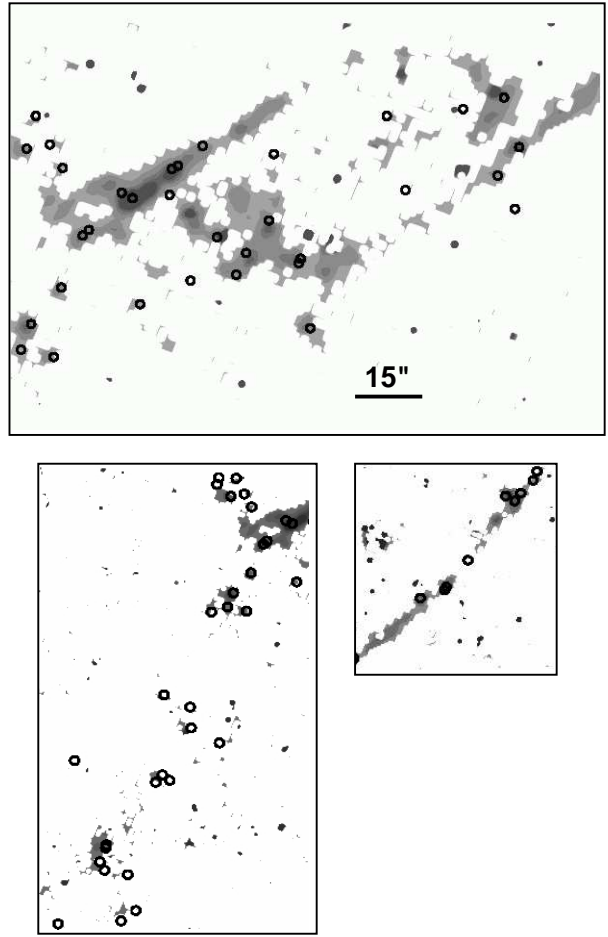}
     \caption{The young ($< 10$ Myr) massive ($\gtrsim 10^{4.5}
         M_{\odot}$) clusters overlayed on H$\alpha$ intensity maps.
     The top panel is the central region of the galaxy (approximately
     region F in Fig.~\ref{regions_dfe}).  The bottom left and right panels
     show the east (regions A, B, and C from Fig.~\ref{regions_abc}) and
     west (region G from Fig.~\ref{regions_gh}) tails respectively.}
         \label{ngc6872-ha}
   \end{figure}

In practice, the construction of a mass distribution of a population of
clusters of many ages is difficult, as over-estimating the age of a
cluster will also cause a significant over-estimate of the mass.  This
is especially the case for clusters with ages $\leq 100$ Myr, where
the mass to light ratio $(M/L)$ of SSP models increases rapidly.  In
principle, if the cluster formation rate is high enough, one can
construct a mass distribution of clusters whose ages are extremely
young, which minimizes the chance of over-estimating their masses.  We
have carried out such an analysis with the sample generated above,
i.e.~those clusters which are spatially coincident with a peak in the
H$\alpha$ intensity and pass our $\chi_{\nu}^2$ criterion.  The
resulting mass distribution is shown as the shaded histogram in
Fig.~\ref{n6872-mass}.  We note that this distribution resembles that
of the entire population, in particular all of the most massive young
clusters are detected in H$\alpha$.  The deviation below $10^{6}
M_{\odot}$ is presumably due to the fainter detection limit in
H$\alpha$ than in the broad band filters.

\section{Specific U-band Luminosity}
\label{specu}
A useful tool to study populations of young clusters and their
environments is the specific {\it U}-band luminosity, first proposed
by Larsen \& Richtler (2000).  It is defined as
\begin{eqnarray}
T_{L}(U) = 100 \times \frac{L_{\rm Clusters}(U)}{L_{\rm Galaxy}(U)}
\end{eqnarray}
where $L_{\rm Cluster}(U)$ is the total amount of light coming from
clusters in the {\it U}-band, and $L_{\rm Galaxy}(U)$ is the total
{\it U}-band light of the galaxy, or of a specific region.  $T_{L}(U)$
has been found to correlate with the area-normalized star formation rate,
$\Sigma_{\rm SFR}$, and to be more or less insensitive to the
low-luminosity tail of the young cluster luminosity function (Larsen
\& Richtler 2000). 

Although the $T_{L}(U)$ relation was defined for an entire cluster
system with respect to the integrated luminosity of the host galaxy,
the concept has been applied to individual cluster complexes 
within a galaxy (Larsen et al. 2002). In the present case we have
sub-divided the galaxy into three distinct regions, the
east tail (roughly regions A, B, and C in Fig.~\ref{image}), the
center (the central part of region F), and the west tail (the west
part of region F and regions G and H).  We then calculated the total
U-band luminosity of each region, ${L_{\rm Galaxy}}$, as well as that
for all point-like sources with $B - V < 0.45$.  The colour cut is an
attempt to homogenize our results with respect to  Larsen \& Richtler
(2000) who use the same cut in order to eliminate foreground stars.

For the Larsen \& Richtler (2000) relation between $T_{L}(U)$ and the
area normalized star formation rate, we used a fit through the data of
Fig. 8 in Larsen (2000), namely
$$ \Sigma_{SFR} [M_{\odot} yr^{-1} kpc^{-2}] = (T_{L}(U)+0.53)/263. $$
Inserting the derived values of $T_{L}(U)$ into the  Larsen \&
Richtler (2000) relation allows us to infer the star formation rates 
for the different regions of the galaxy.

While the  Larsen \& Richtler (2000) relation seems to hold  
for normal galaxies and for galaxies with high star formation rates,
there is considerable scatter in the trend.  Thus, while the absolute
star formation rates estimated in this way are uncertain, the
comparison of the SFR for different regions still allows a relative
interpretation. Table~\ref{n6872-sfr} gives the resultant SFR for each
region.  We see that the star formation rate in the eastern tail is
over twice that of the western tail and more than five times that of
the central region of the galaxy.  We
note that the star formation rate in the central region
of the galaxy is consistent with that
of Mihos et al. (1993), who estimated the SFR $ <
5.7~M_{\odot}~yr^{-1}$ using far infrared colours.

The eastern tail of NGC~6872 seems to dominate the star formation for
the entire
galaxy, while the central region has a rather low SFR.  Mihos et
al. (1993) speculate that the reason for this low SFR is a lack of gas
in the central part of the galaxy.  If most of
the gas was located in the outskirts of the galaxy at the time of the
interaction most of it would be lost into the tails.  This scenario is
consistent with the results of Reif et al. (1982) who did not detect
HI in NGC~6872, as well as with the observations by Horellou \&
Koribalski (2003) which show large amounts of HI inside the tidal
tails.  As can be seen in Figures~\ref{posage} \& ~\ref{ngc6872-ha}, the
majority of the young star clusters are found in the tidal tails or
the outer regions of the disk, consistent with the scenario above.

\begin{table*}[t]
\vskip 1mm
\begin{center}
\begin{tabular}{ccccc}
\hline
Region & $T_{L}^{U}$ & $\Sigma_{SFR} (10^{-2} M_{\odot} yr^{-1}
kpc^{-2})$ & Area ($kpc^{2}$) & SFR ($M_{\odot} yr^{-1}$)\\
\hline

East Tail &  6.6 (1.42)  &    2.7 (0.5) &   605  &    16.5 (3.3) \\
West Tail & 4.8 (1.0)  &   2.1 (0.4) &   375   &   7.6 (1.4) \\
Center  & 2.3 (0.5)   &   1.1 (0.2) &    291   &   3.1 (0.5) \\
\hline
\end{tabular}
\caption{Specific U-band luminosity for the different regions of
NGC~6872 and the implied star formation rates using the Larsen \& 
Richtler (2000) relation.  The numbers in the parentheses represent
the errors of the derived quantities.  The errors were calculated by
comparing the derived values using only clusters with errors in their
colours less than 0.2 with that of using no error criteria.}
\label{n6872-sfr}
\end{center}
\end{table*}

\section{A Scenario for the Formation and Evolution of Star Clusters in
Tidal Tails}
\label{modelcomp}
\subsection{Cluster formation}
The age distribution as a function of position in the galaxy, as seen
in Fig.~\ref{posage}, can be compared with
models of structure formation in the tidal tails, in order to
explore the formation history and triggers of star formation in the
galaxy and tidal tails.

Mihos et al. (1993) have modeled the NGC~6872 interaction in detail,
including a threshold for star formation.  The authors argue that the
coincidence of strong gas compressions with regions of high star
formation rates, is evidence that collisionally induced star formation
is the dominant mode of star formation.  Our result that many of the
YMCs are forming within the regions of strong gas compression and high
star formation rate (see
Fig.~\ref{ngc6872-ha}) agrees with the Mihos et al. (1993) scenario.
Additionally, in both tails we observe a sudden 
drop in $H\alpha$ intensity and density of young clusters directly
outside the regions with high star formation rate along the tails.
After this decrease, we find sporadic single peaks in the $H\alpha$
intensity, coincident with young clusters.  To explain this,
we combine the results from the models of Mihos et al. (1993) with
those of structure formation in the tails (e.g.~Barnes \& Hernquist
1992; Elmegreen et al. 1993).

By combining these models with our observations, we advance the
following scenario.  As the gas gets expelled into the tails, it
collides with other gas clouds, causing an initial burst 
of collisionally induced star formation (Mihos et al. 1993).
Following this initial burst of star/cluster formation the gas is
allowed to cool, as it expands further into the tail.  This explains
the sudden drop in $H\alpha$ intensity along the tails after the large
peak, at the base of the tail (i.e.~the part of the tail closest to
the galaxy).  Due to cooling, the gas it will begin to form denser
clouds and fragment, resembling those found in simulations of the formation
of tidal tails (e.g.~Barnes \& Hernquist 1992 \& Elmegreen et
al. 1993).  This collapse of clouds and structure in the tidal tails by
self-gravity was also suggested by Hibbard \& van Gorkom (1996) to
explain regions of reduced surface brightness in tidal tails.
These clouds, in turn, will be the sites of individual regions of
cluster formation.  Spatial peaks of cluster formation, separated by
large gaps, have also been seen in the tidal tails of the ``Tadpole'' galaxy
(UGC 10214 - Tran et al. 2003).

If this scenario is correct, we may expect older disk stars, that
have been ejected into the tail, to act as seeds for the
gravitational collapse of clouds (Barnes \& Hernquist 1992).  In this
way, the structures formed are expected to be made up of an older disk
population and a younger burst population that forms from the
condensing gas.  Evidence for this scenario has been given by
Weilbacher et al. (2000, 2004), who find that the colours of Tidal Dwarf Galaxy
candidates can be well fit by population synthesis models which contain
both populations. 
As the Tidal Dwarf Galaxy candidates also presumably form in the tidal
tails (Duc \& Mirabel 1998, Weilbacher et al. 2000), it is tempting to
consider a common formation mechanism, 
although this is not necessarily the case.   Optical and near-infrared
spectroscopy/photometry will be necessary to determine if there is a
significant underlying old stellar population within the young
clusters presented here (e.g. Weilbacher et al. 2004).

\subsection{Fate of the Tidal Material}

{\it N}-body simulations of galaxy interactions and the subsequent
formation of tidal tails show that the majority of the material thrown
into the tail will fall back onto the host galaxy within a few 100 Myr
(Hibbard \& Mihos 1995).  This material though, is dissipative gas, while any
non-dissipative structures (e.g.~stars and clusters) will become
ballistic, and in an isolated environment will orbit between the outer
and inner turning points indefinitely (Hibbard \& Mihos 1995).  The
orbits of this material may reach distances of several hundred kiloparsecs with
periods of a few Gyr.   Based on this, we expect much of the gas seen
in the tails of NGC~6872 to fall back onto the main body of the
galaxy, while the YMCs will continue to orbit far outside the galaxy.

The above mentioned simulations were carried out assuming that the galaxy
interaction/merger took place in an isolated environment.  NGC~6872
is, however, a member of the Pavo group, and as such, the tidal
material will experience the potential of the group.  As such, a set of
more realistic simulations may be that of Mihos (2004a,b). In these
{\it N}-body simulations, the author places the galaxy merger (and
hence the tidal tails) in the potential of a rich, Coma-like cluster.
The authors
find that $> 30\%$ of the material thrown into the tidal tails is
stripped from the host galaxy by the cluster potential, into the
intra-cluster medium.  The cluster
potential of the simulated Coma-like cluster ($M_{tot} = 10^{15}
M_{\odot}$) over-estimates the potential
of the much smaller Pavo group by about a factor of 50 to
100\footnote{The mass of the Pavo group was estimated using its
measured temperature of 0.82 keV (Davis, Mulchaey, \& Mushotzky 1999),
which is typical of poor galaxy clusters with masses of a few times
$10^{13} M_{\odot}$ (Mulchaey et al. 1996).}.  Although the Pavo group
is relatively small compared with the simulated Coma-like cluster, the
same mechanism is likely to be at work.  Due to the lower velocities
in small galaxy groups, relative to galaxy clusters, the group
potential may in fact be more efficient in stripping the tidal debris
away from the host galaxy.   In either case, we can expect many of the
star clusters reported here to contribute to the intergalactic
environment of the group. 

As shown in Maraston et al. (2004), massive young star clusters have
parameters (such as size, mass, and $\kappa$ values) that resemble
those of the so-called Ultra-Compact Dwarf
Galaxies (UCDs) discovered in the Fornax cluster (Hilker et al. 1999;
Drinkwater et al. 2000) and
Abel 1689 (Mieske et al. 2004 - though for a discussion on the
relevance of the name see Kissler-Patig 2004).  Despite this similarity, the
prevailing interpretation of the UCDs is that they are the nuclei of
nucleated dwarf galaxies which have had the rest of the galaxy
stripped off due to tidal interactions within the galaxy
cluster, the 'threshing scenario' (Bekki et al. 2003).  On the other
hand, it may be possible that the UCDs are star clusters which have
been formed in the tidal tails of a galaxy merger/interaction.  Due to
the influence of the galaxy cluster potential (see above) many of the
star clusters formed in this way would be expected to be stripped from
the host galaxy, and contribute to a intra-galaxy cluster population
of star clusters.  These clusters would then follow the
velocity/spatial distribution of dwarf galaxies, rather than that of
the globular cluster system of the host galaxy.  Thus, the observed
velocity and spatial distribution of the UCDs in Fornax (Drinkwater et al.~2000) does not rule out the possibility of them being star clusters.

Although NGC~6872 is not currently forming any star clusters with
masses comparable to that of the observed UCDs, larger mergers of
gas-rich galaxies may be able to do so (e.g.~NGC~7252, Maraston et
al. 2004).  This is due to the relation
between the strength of the interaction/merger and the mass of
structures that form in the tidal tails, with stronger encounters
producing larger structures (Elmegreen et al. 1993).

\section{Conclusions}
\label{conclusions}

We have investigated the star cluster population in the tidal tails and
main body of NGC~6872 with $U$, $B$, $V$, and $I$ observations, along with
$H\alpha$ Fabry-Perot observations taken from the
literature (Mihos et al. 1993).  By the use of 
colour-colour diagrams and the {\it 3DEF} method we have 
derived the age distribution of the population along with the
corresponding mass distribution.  We find a rich
population of young ($<$ 10 Myr) massive clusters inside the
tidal tails, which when compared with the dynamical timescale of the
tidal tails themselves ($\sim 145$ Myr) show that the clusters have formed
in the tidal debris.

We find young ($< 10$ Myr) and intermediate ($< 30$ Myr) clusters
spread throughout the tidal tails and main body of the NGC~6872.  Clusters
older than this are
concentrated towards the  main body of the galaxy.  We interpret this as
the cluster 
formation in the tidal tails being delayed by 50--100 Myr with respect
to the event that triggered the formation of the tidal tails themselves.
This will be the case if a significant time delay is required for
clouds to contract under self-gravity in the expanding tidal tails.
Additionally we do not find any strong correlation between the
extinction of the clusters and the HI column density. Young massive
($> 10^{5} M_{\odot}$) clusters are found throughout the 
tidal tails and in the region between the two galaxies.  We find
clusters with masses up to $10^{6.5} - 10^{7} M_{\odot}$ forming due
to the interaction between NGC~6872 and IC~4970.  We may expect even
more massive clusters to form from stronger (e.g.~equal mass)
encounters/mergers (Elmegreen et al. 1993).

The mass function of the population is very similar to that of
other young populations in various environments (e.g.~galaxy mergers,
nuclear starbursts, normal spirals), being well fit by a 
power-law of the form $N(m) \propto m^{-\alpha}$ where $\alpha = 1.85
\pm 0.11$.  Due to the intrinsic low spatial resolution of ground
based images, we may have slightly underestimated $\alpha$ by mistaking
a group of young clusters as one single cluster.  Thus, we expect that
$\alpha$ would be somewhat larger on higher resolution images.

Using the specific U-band luminosity $T_{L}(U)$ of the tidal tails and the main
body of NGC~6872, and combining this result with the empirical
$T_{L}(U)$ vs. $\Sigma_{\rm SFR}$ relationship (Larsen \& Richter
2000), we have estimated the star formation rate for each area.  While
the actual numbers derived in this way are relatively uncertain, the
relative star
formation rates should hold.  Using this method we find that the east tidal
tail has the highest star formation rate ($\sim 16.5 M_{\odot}
yr^{-1}$) at present, with the west tail having about half of this
value, and the star formation rate in the central region of the galaxy
being about five times lower than the eastern tail.

By comparing the spatial position of the youngest clusters with
$H\alpha$ observations of NGC~6872, we can get an independent
constraint on the ages of the
clusters and eliminate foreground stellar contamination.  We
find that the most massive ($> 10^5 M_{\odot}$) and youngest
($< 10$ Myr) clusters coincide with peaks in the
$H\alpha$ intensity supporting our age and the resulting mass
estimates based on 
colours.  The tidal tails show prominent large gaps in the
$H\alpha$ intensity which coincide with large gaps in the cluster
population.  This is 
consistent with the scenario of Hibbard and van Gorkom (1996) who
propose that clouds contract under self-gravity inside the tails,
leaving areas of low surface brightness between the clouds.   If we
apply our analysis to only those clusters which coincide with peaks in
the $H\alpha$ emission we find that these clusters account for all of
the most massive young clusters in our sample.  We find that there are
many massive ($> 10^{5} M_{\odot}$) clusters present in the tidal
tails and the outer parts of the galactic disk. 

Finally, we have qualitatively combined our observations with detailed
models of structure formation in tidal tails as well as their
evolution.  These models conclude that a large fraction ($\sim 30$\%) of the
material in the tidal tails, and a possibly larger fraction of the
material in stars and star clusters, will be expelled into the
intra-galaxy cluster environment.  Although the Pavo group is a factor
of 50--100 smaller than the simulated Coma-like cluster, the smaller
velocities of the galaxies may make tidal stripping even more
efficient.  Since the same stripping mechanism is expected to be
operating on both the galaxy cluster and galaxy group scales, NGC~6872 can
be used as an interesting test case.  We have shown that massive
clusters can form in the tidal tails, far from the main body of the
galaxy.  This tidal debris will be stripped
into the intra-cluster environment, and form a population of star
clusters that follow the galaxy cluster potential in terms of
velocities and spatial distribution.  If the central cD galaxies of
Fornax-like galaxy clusters were formed by the merging of galaxies
which resulted in tidal tails, we expect a population of star clusters
in the intra-cluster environment.  The most massive clusters are
expected to have similar properties to the observed ultra-compact
dwarf galaxies found in Fornax and Abel 1689.

\begin{table*}[!p]
{\scriptsize
\parbox[b]{18cm}{
\centering
\begin{tabular}{c c c c c c c c c c c c c c c c}    
\hline\hline
\noalign{\smallskip}
id & x & y & U$^{a}$ & dU & B$^{a}$ & dB & V$^{a}$ & dV & I$^{a}$ & dI & $\chi^{2}_{\nu}$$^{b}$ & A$_{V}$$^{c}$ & log (Age/yr)$^{c}$ & log Mass/M$_{\odot}$$^{c}$ & M$_V$$^{d}$ \\ 
\hline\hline
\noalign{\smallskip}  

  2  &    120.6  &    164.0  &    23.25  &     0.17  &    23.05  &     0.04  &    22.27  &     0.03  &    21.37  &     0.04  &     2.00  &     0.00  &     9.48  &     6.89  &   -11.63  \\
  6  &    189.0  &    178.2  &    22.55  &     0.11  &    23.34  &     0.07  &    23.58  &     0.09  &    22.90  &     0.10  &     7.60  &     0.19  &     7.20  &     5.11  &   -10.51  \\
 10  &    215.5  &    197.7  &    22.05  &     0.09  &    22.58  &     0.05  &    22.73  &     0.07  &    22.50  &     0.09  &     1.80  &     0.06  &     6.64  &     4.59  &   -11.23  \\
 14  &     40.4  &    236.0  &    20.94  &     0.03  &    20.89  &     0.01  &    20.46  &     0.01  &    19.77  &     0.03  &     0.80  &     0.00  &     8.96  &     7.20  &   -13.44  \\
 16  &     71.0  &    240.8  &    21.83  &     0.07  &    22.21  &     0.03  &    22.28  &     0.04  &    21.96  &     0.06  &     6.80  &     0.25  &     6.62  &     4.79  &   -11.87  \\
 19  &    201.0  &    265.4  &    22.09  &     0.08  &    22.54  &     0.03  &    22.61  &     0.04  &    22.13  &     0.06  &     9.20  &     0.19  &     6.88  &     4.95  &   -11.48  \\
 20  &    159.4  &    273.8  &    22.75  &     0.17  &    23.14  &     0.08  &    23.07  &     0.09  &    22.15  &     0.07  &     3.10  &     0.74  &     7.30  &     5.52  &   -11.57  \\
\noalign{\smallskip}
\hline
\end{tabular}
\begin{list}{}{}
\item[$^{\mathrm{a}}$] Magnitudes have only been corrected for foreground extinction.
\item[$^{\mathrm{b}}$] The reduced $\chi^{2}$ value for the best fitting model.
\item[$^{\mathrm{c}}$] Derived extinction, age, and mass using the {\it 3DEF} method.  Sources with ages greater than 1 Gyr are most likely foreground stars.
\item[$^{\mathrm{d}}$] Absolute magnitue of the source, correcting for the derived extinction.

\end{list}
\caption[tails1]{Photometric and derived properties of candidate
  clusters in NGC 6872.  The complete list is available in electronic form
at the CDS via anonymous ftp to cdsarc.u-strasbg.fr (130.79.128.5)
or via http://cdsweb.u-strasbg.fr/cgi-bin/qcat?J/A+A/\\
 }
\label{sources}
}
}
\end{table*}

\begin{acknowledgements}
We thank Chris Mihos for
providing his $H\alpha$ data in electronic form, as well as for
many useful discussions.  We also thank S\o ren Larsen for useful
discussions concerning the specific U-band luminosity, and Henny
Lamers for comments on the manuscript.  Cees Bassa is gratefully acknowledged for providing the astrometry.  Finally, we thank the
VLT staff for their excellent help during the observations.
	
\end{acknowledgements}

\bibliographystyle{alpha}
\bibliography{../bib/astroph.bib,../bib/phd.bib,../bib/mark.bib}

\end{document}